# Light control of the diffusion coefficient of active fluids

**Thomas Vourc'h**
Laboratoire AstroParticules et Cosmologie, CNRS, Université de Paris
5 rue Thomas Mann 75013 Paris, France
Present address: Physico-Chimie Curie, CNRS, Institut Curie, Paris Sciences et Lettres, 11 rue Pierre et Marie Curie 75005 Paris, France
thomas.vourch@curie.fr

**Julien Léopoldès**
ESPCI Paris, PSL Research University, CNRS, Institut Langevin
1 rue Jussieu, F-75005 Paris, France
Université Paris-Est Marne-la-Vallée
5 Bd Descartes, Champs sur Marne, Marne-la-Vallée cedex 2, France
julien.leopoldes@u-pem.fr

**Hassan Peerhossaini[1]**
Laboratoire AstroParticules et Cosmologie, CNRS, Université de Paris
5 rue Thomas Mann 75013 Paris, France
Mechanics of Active Fluids and Bacterial Physics Laboratory,
Department of Civil and Environmental Engineering, Department of Mechanical and Materials Engineering,
University of Western Ontario, London, ON, Canada
hpeerhos@uwo.ca

**ABSTRACT**
*Active fluids refer to the fluids that contain self-propelled particles such as bacteria or micro-algae, whose properties differ fundamentally from the passive fluids. Such particles often exhibit an intermittent motion, with high-motility "run" periods broken by low-motility "tumble" periods. The average motion can be modified with external stresses, such as nutrient or light gradients, leading to a directed movement called chemotaxis and phototaxis, respectively.*
*Using cyanobacterium Synechocystis sp.PCC 6803, a model micro-organism to study photosynthesis, we track the bacterial response to light stimuli, under isotropic and non-isotropic (directed) conditions. In particular, we investigate how the intermittent motility is influenced by illumination.*
*We find that just after a rise in light intensity, the probability to be in the run state increases. This feature vanishes after a typical characteristic time of about 1 hour, when initial probability is recovered. Our results are well described by a mathematical model based on the linear response theory.*
*When the perturbation is anisotropic, we observe a collective motion toward the light source (phototaxis). We show that the bias emerges due to more frequent runs in the direction of the light, whereas the run durations are longer whatever the direction.*

---

[1] hpeerhos@uwo.ca

1. **INTRODUCTION**

Contrary to the conventional fluid flows in which one needs gradients of pressure, velocity and temperature to break equilibrium and drive the flow, in active fluids, biological cells which are the microstructural elements of the fluid have their own molecular motor that can activate appendixes such as flagella or pili for driving the flow. Despite our current understanding of active fluid mechanics, a significant knowledge gap exists in the quantitative understanding and modelling of the mechanisms underlying many manifestations of active fluid.

The energy needed to trigger molecular motors and perform other metabolic functions can be found in chemical nutrients or in the surrounding light for phototroph organisms. Hence, these microorganisms tend to find the better conditions for their growth: active fluids have the capacity of analysis and action, though rudimentary.

Gradients of nutrients can thus result in collective movements toward the nutrient source, according to a mechanism called chemotaxis [1,2]. The application of a light gradient to other strains can lead to a similar behaviour called phototaxis [3]. This phenomenon, coupled to bacterial interactions, can generate mesoscopic fingering instabilities [4]. The control of the motility of bacteria with light has been the subject of many recent studies [5–7].

In this paper, we use the cyanobacterium *Synechocystis* sp. PCC 6803 as a model microorganism. This is a unicellular prokaryote, whose genome has been completely determined [8], and which is a typical microorganism for the investigation of photosynthesis [9]. We have previously characterized the intermittent motility of these bacteria, based on the alternation of "run" periods during which they move, and "tumble" periods which consist of localized motion [10,11]. We have also studied the effects of hydrodynamic stress on the growth of *Synechocystis* sp. PCC 6803 [12]. Of special interest is the fact that despite its biological simplicity this microorganism is able to adapt itself to the conditions of the surface on which it diffuses [11].

*Synechocystis* exhibits an intermittent motility without external perturbation. However, the question rises as to whether the bacteria adapt their motility to changes in isotropic light intensity? Especially, are the characteristic times for intermittence involved in this response? This study is aimed at describing how light conditions can influence the diffusion of individual cells. This could bring new insights in the control of the motion of an active fluid composed of living micro-organisms, both in terms of diffusion and of directionality.

In this work, we first investigate the influence of the light intensity on the bacterium motility in an isotropic assay where bacteria undergo steps of isotropic light flux. We model analytically the results in the framework of what has been proposed for bacterial chemotaxis [13–15]. We complement this study with a phototactic experiment aimed at uncoupling the effect of light intensity from that of the light direction. Taken together, these studies describe how light and light direction can be used to control active fluids.

## 2. Materials and methods

### 2.1 Microorganism and culture conditions

Experiments are carried out with suspensions of the unicellular cyanobacterium *Synechocystis* sp. PCC 6803, a model of environmentally important photosynthetic prokaryote. Typical Reynolds number for such microorganisms (density=$1.1\ 10^3$ kg.m$^{-3}$, diameter=3 micrometers) are indeed really low, of the order of magnitude of $10^{-6}$. The *Synechocystis* cells used here display a spherical shape when they are not dividing, so the diffusion coefficient remains a scalar. We have opted to use a small magnification to ensure the statistical relevance of our study, meanwhile we note that using higher magnifications could be used to characterize the diffusion of dividing cells and hence observe their orientation in relation to the flow.

The *Synechocystis* sp. PCC 6803 strain used here is collected in the Pasteur Culture Collection of cyanobacteria (Paris, France). The strains are routinely cultured in the BG11 standard mineral medium, and sub-cultivated by diluting 3 mL of a mother culture in 47 mL of fresh BG11. The suspensions are stirred by a magnetic agitator operating at 360 rotations per minute in a clean room at 20 °C. They are placed under white light intensity of 1.3 W.m$^{-2}$ for 7 days followed by 24 hours dark and subsequent 2 hours light before running the experiments. At this stage, the concentration of cyanobacteria is approximately $2.10^7$ cells per mL. The suspensions are diluted 10 fold in fresh BG11 before introduction in the measurement microchips.

**2.2 Experimental conditions**

A droplet of cellular suspension is placed in the cavity of a microscope glass slide (BRANDT, 26×76 mm). The cavity is then closed with a glass coverslip (Menzel-Gläser, 22×22 mm) and sealed with elastomer to avoid evaporation. The obtained observation cell is then turned over so that we observe the displacement of the bacteria on the glass coverslip, and placed on the stage of an inverted optical microscope coupled to a CCD camera. Detailed description of the experimental set-up and procedures are given in [11].

**2.3 Imaging techniques**

Images are acquired with a CCD camera equipped with a Nikon TU Plant 10X objective, with a resolution of 1.6 pixels per micron. The observed area is illuminated with white light passing through an optical fiber and homogenized by an optic tube.

**2.4 Control of isotropic light**

The intensity provided by the incident white light used for imaging with the microscope can be tuned. This changes the luminous flux to which the bacteria are exposed. The control of isotropic light is ensured manually with a potentiometer on the light source. Every hour, the position of the potentiometer is manually changed in order to raise the desired light flux. We have previously measured the light flux and identified the positions of the potentiometer corresponding to the magnitude of the desired light flux. We have opted to set two light intensities: a "normal" intensity of $I_0$=465 lux and a "strong" intensity of 665 lux. The intensity difference between the two periods is noted by $\Delta I$, so that the "strong" intensity is $I_0+\Delta I$.

Experiments start by setting the light intensity at $I_0$ for two hours so that the bacteria reach their motility plateau [11]. The intensity is then increased to the higher value for one hour, which defines a step. Then it is reduced to $I_0$ for one hour. The succession of increasing /decreasing step defines a light cycle.

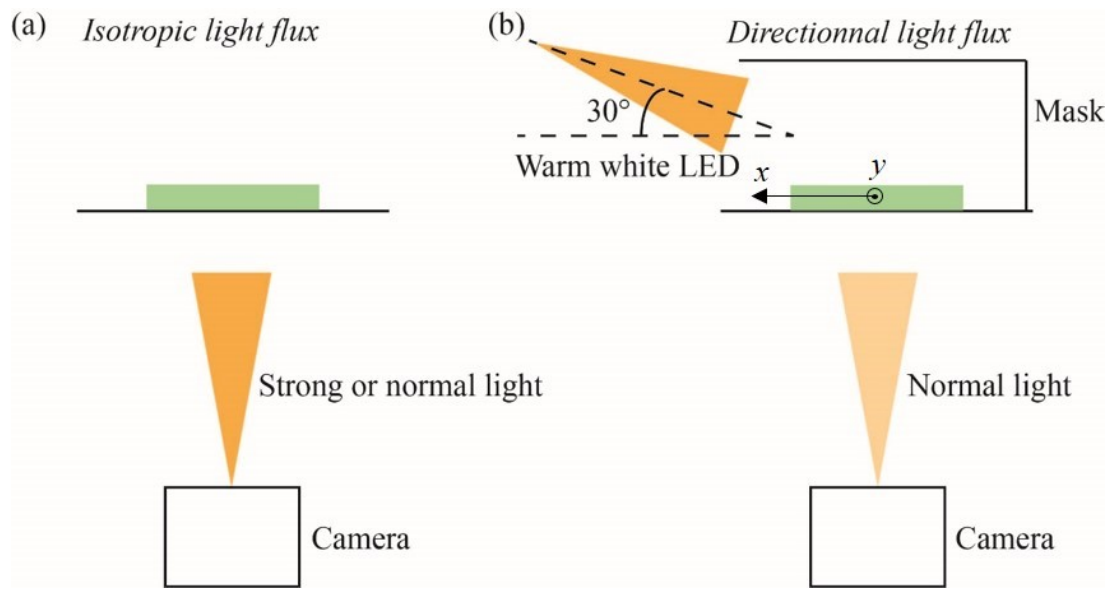

**FIGURE 1:** Schematic view of the experimental setups used in this study. (a) Response to an isotropic perturbation, (b) response to an anisotropic (directional) perturbation.

### 2.5 Control of anisotropic light

The light intensity gradient is obtained with a Farnell LED (5 mm, 4000K, 4.7 cd), used with a power supply that enables us to tune its intensity. An in-house light orientating plateau has been designed and constructed to direct the incident light with an angle of 30° towards the bacterial sample. A cover mask has also been added in order to minimize the unwanted illumination from surrounding lights. A summary of the different light conditions protocol used here is provided in FIGURE 1.

As for the isotropic light conditions, the light intensity is set to $I_0$ for two hours. Then the LED is turned on for one hour, and turned off for the following one, which defines a two-hour light cycle.

For the experiment under isotropic (directional) light, six cycles of two hours each (one hour of high intensity followed by one hour of normal intensity) are analyzed. The phototaxis assay describes the results obtained for three light cycles. The cell population has been renewed after each assay.

### 2.6 Post-processing

Video recordings have been post-processed with the software ImageJ. Acquisition frequency is set to 1 frame per second. Bacteria positions and experimental trajectories are recorded using MATLAB code [16] customized for our purpose.

### 3 Isotropic perturbations

### 3.1 Experimental results

Once the bacteria (in suspension) have been placed in the measurement cell, they are let free to sediment on the lower surface of the cell and diffuse for two hours, so that their diffusion coefficient reaches a plateau value, as is explained in [11]. Then, light intensity is increased for one hour and we record the changes in the motility that occurs after this perturbation. After one hour of high intensity, the light is reduced again to I₀ for one hour, and we record the response to this descending step. A description of the illumination profile is given in FIGURE 2. This experiment has been repeated three times, so that we have 6 cycles.

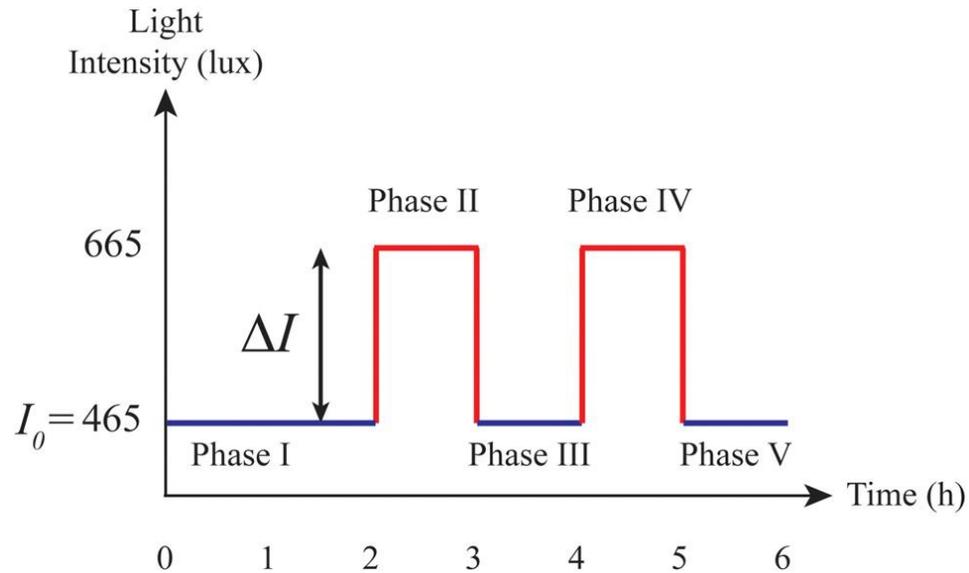

**FIGURE 2:** Temporal profile of the light intensity protocol used in the isotropic illumination experiment. During Phase I, light intensity is set to $I_0$ for two hours. It is then raised to $I_0+\Delta I$ for one hour (Phase II), and reduced to $I_0$ again for the subsequent hour (Phase III). The latter cycle is repeated once (Phases IV-V).

As a marker for motility, we compute a temporal Mean-Squared-Displacement for a time interval Δ, averaged over all the trajectories and all the displacements recorded along a temporal window centered at around time t. This can be written as:

$$MSD(t,\Delta) = \langle (X(\tau + \Delta) - X(\tau))^2 \rangle_t \quad (1)$$

where $X(\tau)$ is the position of a bacterium at time $\tau$ and the brackets $\langle . \rangle_t$ denote an average over all trajectories and all times $\tau$ between the times $t - \delta$ and $t + \delta$, with $\delta$=100s.

We find Fickian dynamics that enables us to define a time-dependent diffusion coefficient, defined as:

$$D(t) = \frac{MSD(t,\Delta)}{4\Delta} \quad (2)$$

More details of the computation can be found in [11]. Temporal evolutions of $D(t)$ for two representative light cycles are given in FIGURE 3. Values are normalized by $D_\infty$, the final (asymptotic) value of the diffusion coefficient. Once the luminous flux has been increased, the value of $D$ steps up and reaches a maximum after about 600 seconds, before relaxing to its initial value. When a "normal" intensity is recovered, which corresponds to a "negative" step, we obtain a symmetrical behavior; motility decreases before reaching the plateau value $D_\infty$.

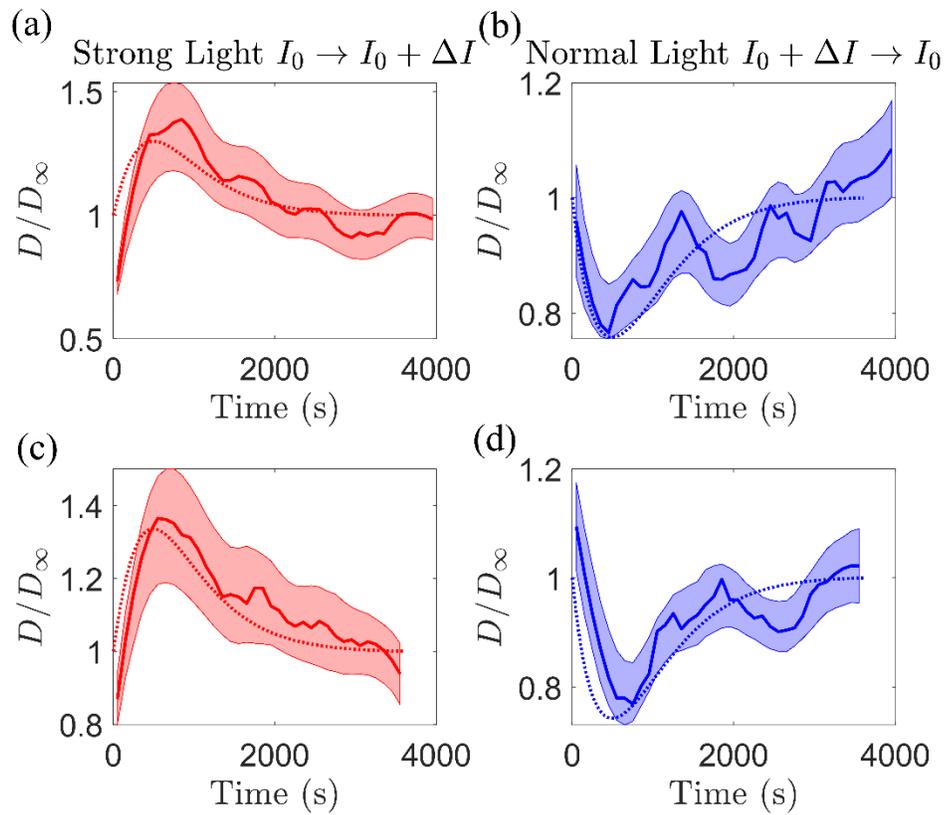

**FIGURE 3:** Temporal evolution of the diffusion coefficient normalized by their final value. Plain line: experimental results, and dotted line: analytical formula (EQ. 13). Colored areas indicate +/- standard deviation.

As several other bacteria, *Synechocystis* displays an intermittent motility, with "run" periods of directed motion and "tumble" periods during which bacteria are non motile [1,17]. The variation of the diffusion coefficient could then arise from variations of the characteristic times for both periods.

We have then computed the characteristic times corresponding to both periods, $\tau_{run}$ and $\tau_{tumble}$ respectively, according to the procedure described in [11]. Briefly, we compute the bacterium displacement in short time intervals along the whole duration of a trajectory. If this displacement is above a certain threshold distance, then the time interval is considered as a "run", otherwise it is a "tumble". A typical trajectory divided into run and tumble periods is shown in FIGURE 4.

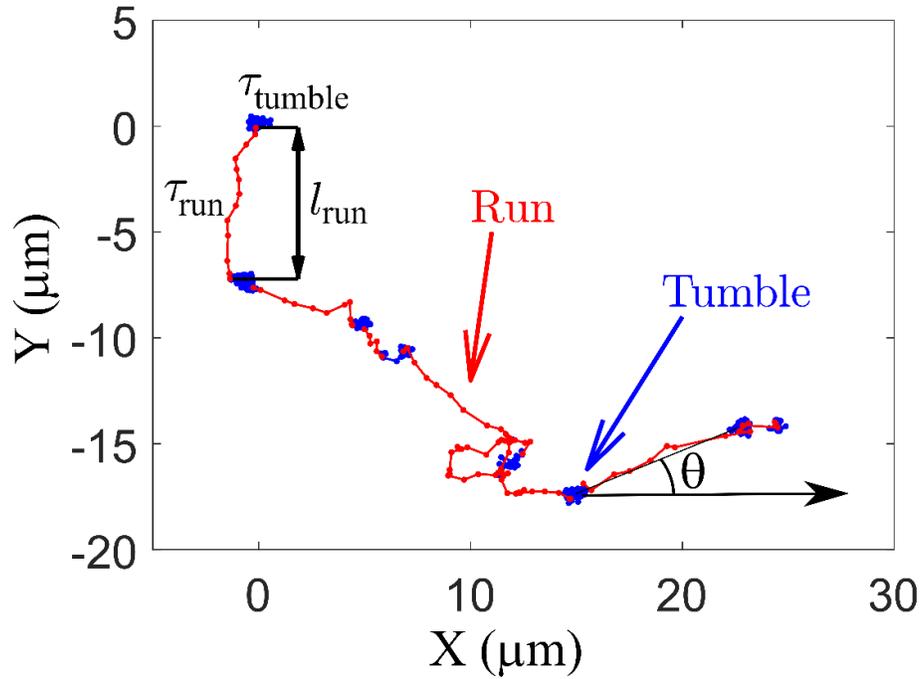

**FIGURE 4**: Trajectory of a bacterial cell (668 s). Runs correspond to red lines and tumble to blue dots. For the phototaxis assay, we define the angle θ between the direction of the run and the direction towards the light source, which is indicated by the arrow.

The temporal evolution of the inverse run and tumble durations are given in FIGURE 5. Data are normalized by their final value $\bar{\tau}_{run}$ and $\bar{\tau}_{tumble}$ and averaged over four light cycles each. After the increase of the light intensity, the inverse run time reduces and the inverse tumble time increases, highlighting a higher propensity to be in a "run" period. This feature slowly vanishes after the peak, so that the initial values are recovered after one hour. FIGURE 5 also shows that the evolutions of the inverse run and tumble times are symmetrical after a decrease of the light intensity.

Journal of Fluids Engineering

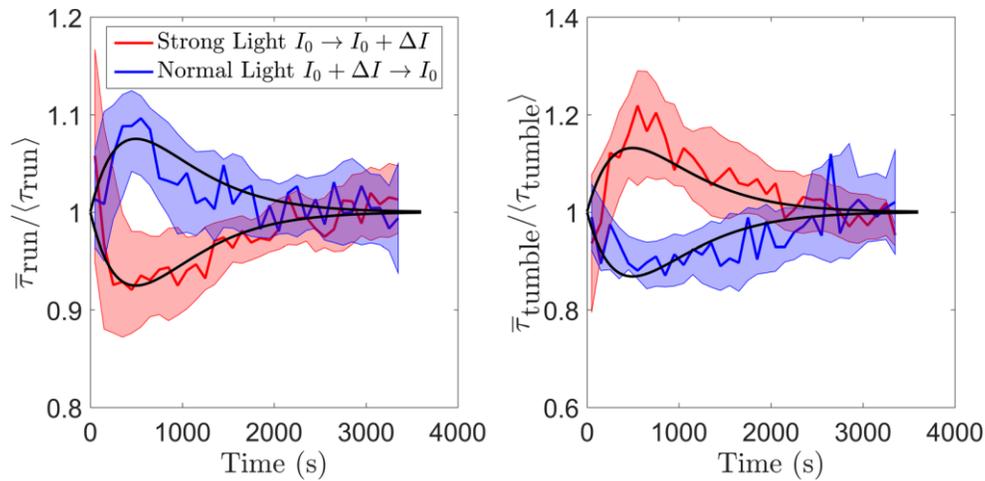

**FIGURE 5:** Temporal evolution of the inverse (a) run and (b) tumble times, rescaled by their final value. Black Plain line: model described by EQ. (7-8). Colored areas indicate +/- standard deviation.

### 3.2 Analytical description

This ability to change the main feature of their motility suggests that *Synechocystis* cells adapt to changing conditions. Such a property has also been disclosed for the chemotactic motility of *Escherichia Coli* bacteria, flagellar micro-organisms responding to nutrient concentration perturbations [18]. Experimental results of this seminal work have given rise to mathematical modelling in the framework of the linear response theory [13,15,19], from which we inspire to model our own experiments.

This approach writes the output of a system $S(t)$ submitted to a small external perturbation $F(t)$ as the convolution of this input with a response function $R(t)$ that is characteristic to the system, according to the following equation:

$$S(t) = \int_{-\infty}^{t} R(t-t')F(t')dt' \qquad (3)$$

This general formula, adapted by De Gennes [13] for the characterization of run and tumble times, provides:

$$\frac{1}{\langle \tau_{\text{run}} \rangle(t)} = \frac{1}{\bar{\tau}_{\text{run}}} \left[ 1 - \int_{-\infty}^{t} R_{\text{run}}(t-t')c(t')dt' \right] \qquad (4)$$

$$\frac{1}{\langle \tau_{\text{tumble}} \rangle(t)} = \frac{1}{\bar{\tau}_{\text{tumble}}} \left[ 1 - \int_{-\infty}^{t} R_{\text{tumble}}(t-t')c(t')dt' \right] \qquad (5)$$

where $R_{run}$ and $R_{tumble}$ stand for the response function related to the two periods of the motility, and $c(t) \leftrightarrow \pm \Delta I$ is the variation of light intensity, $+\Delta I$ when intensity is increased and $-\Delta I$ when it is decreased. Equations (4) and (5) correspond to the general formula given by Equation (3) when defining the output by $S(t) = 1 - \frac{\bar{\tau}_{run}}{\langle \tau_{run} \rangle(t)}$ (same with tumble times). In FIGURE 5, symmetry of the results with respect to the axis $y = 1$ is then consistent with the properties of the linear response theory, which predicts that the response to a $+F$ perturbation is the opposite of the one obtained with a $-F$ perturbation.

Our experimental data are too noisy to enable determining an accurate response function of unknown shape. For this purpose, we adjust the experimental curves in FIGURE 5 with the form proposed by [15,19] describing the chemotaxis of *E. coli* bacteria:

$$R_{\text{run}}(t) = W_{\text{run}}(1 - \frac{\lambda t}{2} - \frac{\lambda^2 t^2}{4})e^{-\lambda t} \qquad (6)$$

where $W_{run}$ is a gain factor and $\lambda$ the inverse of a relaxation time that describes the time needed to recover initial values after the perturbation. To describe the evolution of the tumble times, we also use Equation (6) with the same value of $\lambda$ and a gain $W_{tumble}$, and we note that the evolution of $R_{tumble}(t)$ should be opposite to $R_{run}(t)$. Integrating according to Equations (4) and (5) gives:

$$\frac{\bar{\tau}_{run}}{\langle \tau_{\text{run}} \rangle(t)} = 1 - W_{\text{run}} \Delta I (t + \frac{\lambda t^2}{4})e^{-\lambda t} \qquad (7)$$

$$\frac{\bar{\tau}_{tumble}}{\langle \tau_{\text{tumble}} \rangle(t)} = 1 + W_{\text{tumble}} \Delta I (t + \frac{\lambda t^2}{4})e^{-\lambda t} \qquad (8)$$

with $\Delta I = \pm 200 \, lux$, we find that experimental data shown in FIGURE 5 are well fitted with $\lambda^{-1} = 400 \, s^{-1}$, $W_{\text{run}} = 4.10^{-4} \, lux^{-1}.s^{-1}$ and $W_{\text{tumble}} = 7.10^{-4} \, lux^{-1}.s^{-1}$. The overall duration of a run period and its subsequent tumble period is approximately 80s. The value we obtain for $\lambda^{-1}$ then corresponds to several run/tumble cycles. We note that in the analytical approaches [15,19], the value of $\lambda^{-1}$ is shorter (about 1 s), but also corresponds to the order of magnitude between two run periods. Hence the adaptation of *Synechocystis* cells to changes in light intensity of the environment can be described with the same function that has been proposed for the chemotaxis of another bacterial strain.

Journal of Fluids Engineering### 3.2 Evolution of the diffusion coefficient

We have previously described the diffusion coefficient of *Synechocystis* bacteria in terms of intermittency by the formula [11]:

$$D \sim \frac{1}{4} \bar{V}_m^{\,2} \frac{\langle \tau_{run}^2 \rangle}{\langle \tau \rangle} \quad (9)$$

where $\bar{V}_m^{\,2}$ is the average run speed and $\langle \tau \rangle$ the average time between two successive run periods. We find that $\bar{V}_m$ is constant and equals to $0.44 \; \mu m.s^{-1}$. We approximate $\langle \tau \rangle$ as $\langle \tau_{run} \rangle + \langle \tau_{tumble} \rangle$. Besides, $\langle \tau_{run}^2 \rangle$ and $\langle \tau_{run} \rangle^2$ are related by the variance of the run times distribution:

$$var(\tau_{run}) = \langle \tau_{run}^2 \rangle - \langle \tau_{run} \rangle^2 \quad (10)$$

By studying empirically this variance (data are displayed in Appendix A), we find that for both negative and positive steps, one can write:

$$var(\tau_{run}) = a \langle \tau_{run}^2 \rangle + b \quad (11)$$

with $a = 0.6$ and $b = -16s^2$. Using the definition of the variance (Eq. 10), this leads to:

FE-19-1649 – T.Vourc'h, J. Léopoldès, H. Peerhossaini     15

$$\langle \tau_{\text{run}}{}^2 \rangle = \frac{b + \langle \tau_{\text{run}} \rangle^2}{1-a} \tag{12}$$

For clarity, we define $f_{\text{run}} = W_{\text{run}} \Delta I \left(t + \frac{\lambda t^2}{4}\right) e^{-\lambda t}$ and $f_{\text{tumble}}(t) = W_{\text{tumble}} \Delta I \left(t + \frac{\lambda t^2}{4}\right) e^{-\lambda t}$. Combining Equations (7), (8), (9) and (12) provides the following expression for the diffusion coefficient:

$$D = \frac{\bar{V}_m{}^2}{4(1-a)} \frac{b + \left(\frac{\bar{\tau}_{\text{run}}}{1-f_{\text{run}}}\right)^2}{\left(\frac{\bar{\tau}_{\text{run}}}{1-f_{\text{run}}}\right) + \left(\frac{\bar{\tau}_{\text{tumble}}}{1+f_{\text{tumble}}}\right)} \tag{13}$$

This expression, renormalized by the final values of the diffusion coefficient, is compared to experimental results after different steps of light intensity variation in FIGURE 3. It captures both the initial variation of the diffusion coefficient and its relaxation to its initial value. This analytical approach allows to show how bacteria can adapt to light changes by triggering the characteristic times of their intermittent motility.

These experiments carried out under isotropic conditions show that light intensity influences bacterial motion but do not address the issue of phototaxis, where not only the dynamics, but also the directionality of light, are involved. One thus may wonder how an anisotropic perturbation will affect the results.

## 4   Directional perturbations

### 4.1 Instantaneous phototaxis

The previous part described the adaptation process of *Synechocystis* to an isotropic perturbation. We now study the case of a directional perturbation, where both flux intensity and its direction vary. This perturbation is generated by a warm white LED (see section 1.5 for further details), as is shown in the schematic view of the experimental setup in FIGURE 1(b).

The temporal protocol of the perturbation is similar to the one shown in FIGURE 2: after a two-hour period necessary for the bacteria to sediment and adopt a constant motility, the LED is turned on for one hour, and stopped for the subsequent hour. This alternation defines a light cycle, which is repeated three times. For the following, we define $x$ as the longitudinal axis corresponding to the light direction, with $x > 0$ pointing toward the light source, and $y$ the transverse direction. $x$ and $y$ axes are shown in FIGURE 1(b).

*Synechocystis* cells have been reported to undergo a biased motility under directional illumination [20], which is the definition of phototaxis. We first verify the existence of this phototaxis by computing the average velocities $V_x$ and $V_y$ along $x$ and $y$ directions respectively after the perturbation light source has been turned on or turned off.

The speeds at a time $t$ since the beginning of the perturbation are computed over all the trajectories of duration $\delta = 50\ s$ detected between the time $t$ and $t + \delta$, according to:

$$\begin{cases} V_x = \langle \frac{x(t+\delta)-x(t)}{\delta} \rangle \\ V_y = \langle \frac{y(t+\delta)-y(t)}{\delta} \rangle \end{cases} \qquad (14)$$

Results are shown in FIGURE 6: after the LED has been turned on, $V_x$ rises quickly and eventually reaches a nonzero plateau value. Average velocity in the $y$ direction $V_y$ remains null. This observation implies that bacteria have a propensity to move towards $x > 0$, i.e. towards the light source, while no bias emerges in the transverse direction. When the illumination is reset to an isotropic situation, both time averaged velocities $V_x$ and $V_y$ tend to zero, which implies that motility is then random. We note that at short times after the shutdown of the LED, $V_x$ is slightly negative, but we have no clue to explain this feature.

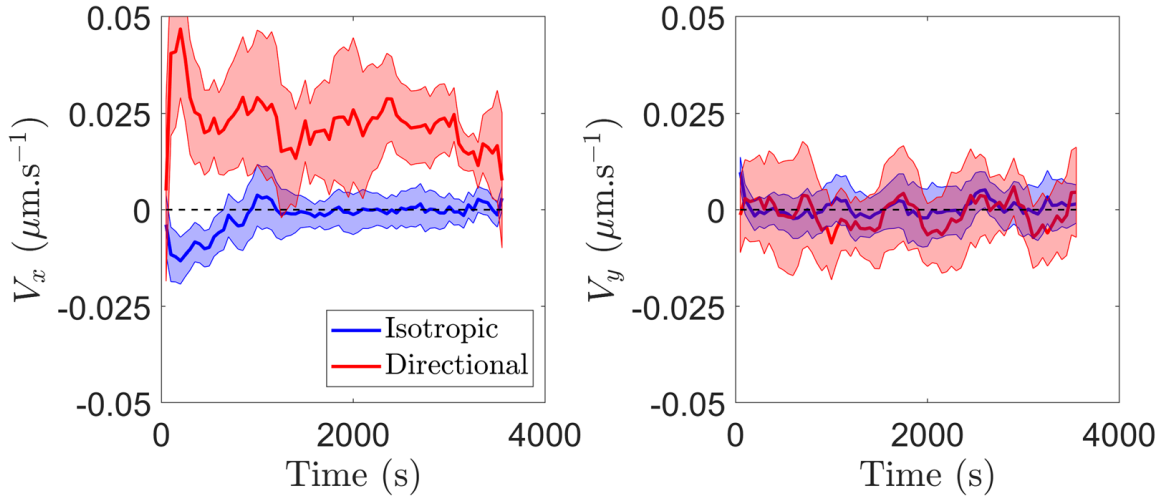

**FIGURE 6:** Average velocity along the longitudinal and transverse directions as a function of the time elapsed since the conditions have changed. Red: After the LED has been turned on (anisotropic illumination) and blue: after the LED has been turned off

(isotropic illumination). Black dashed line indicates a zero velocity corresponding to an absence of biased movement. Colored areas indicate +/- standard deviation.

**4.2 Bias is related to run direction**

We further investigate to better understand the role of the distinct components (run and tumble) of the bacterial movement in the emergence of the bias. We first examine whether the observed bias is caused by directed run periods. We define $\overrightarrow{l_{run}}$ the displacement being performed during a run period, and compute the average projection of this displacement along the $x$ and $y$ axes:

$$\begin{cases} l_x(t) = \langle \overrightarrow{l_{run}} \cdot \vec{x} \rangle \\ l_y(t) = \langle \overrightarrow{l_{run}} \cdot \vec{y} \rangle \end{cases} \quad (15)$$

where the brackets denote an average over all recorded runs between the times $t$ and $t + \delta$, with $\delta = 100s$, $t = 0$ being the instant at which the light conditions are changed.

Temporal evolutions of $l_x$ and $l_y$ are shown in FIGURE 7. Without directional light (phase I), both average projections are zero, suggesting that runs' directions are random. After the LED is turned on (phases II, IV and VI), $l_x$ reaches quickly a positive value, showing that the runs promote bacterial displacement towards the light source; meanwhile $l_y$ keeps its zero value. When the luminous flux is isotropic again (phases III and V), both projections are null. Hence, the nonzero average displacement during run periods is responsible for the bias of the overall motility.

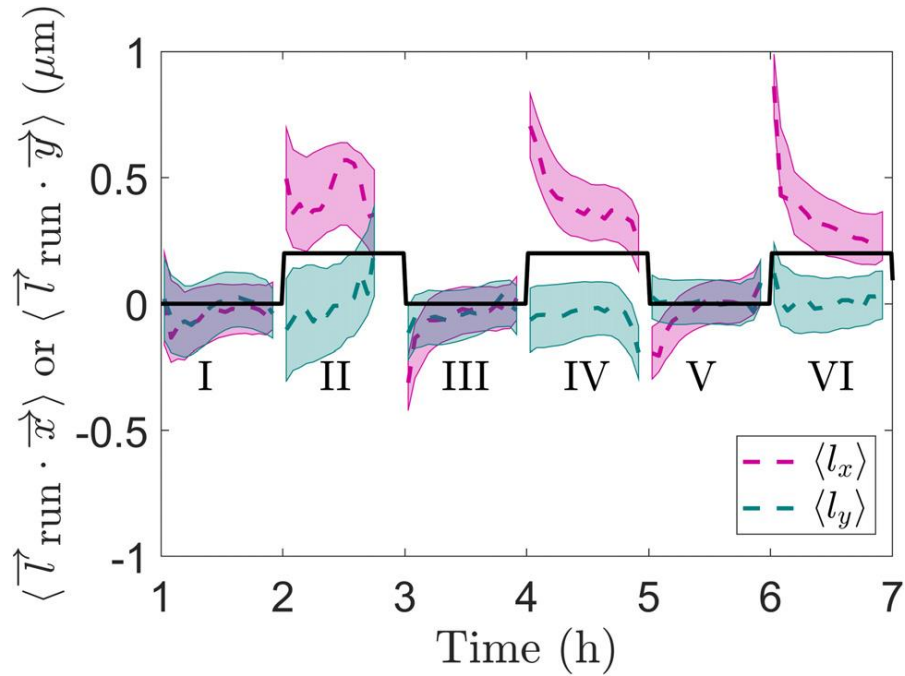

**FIGURE 7:** Temporal evolution of the averaged projection during runs periods along longitudinal $x$ and transverse $y$ directions. Plain black line shows the state of the LED (turned off when it is 0, turned on elsewhere). Colored areas indicate +/- standard error.

**4.3 Bias emerges from runs occurrences, not runs durations**

In this section we focus on the emergence of the bias by studying not only the averaged run statistics but also their distribution according to their corresponding direction. We note $\theta$ the angle between the run displacement and the direction toward the light source, as defined in FIGURE 3 ($\theta = 0$ indicating that the displacement during the considered run is in the direction of the light source). We define 20 segments for $\theta$

Journal of Fluids Engineeringranging from 0 to 360°, and compute for each segment the mean run durations and the occurrence of runs in this section.

FIGURE 8(a) displays the average run durations according to their direction, when the LED is turned on and turned off. For both illumination conditions, run durations are the same regardless their direction (it can be seen from the fact that the resulting plot is a circle in our representation). Meanwhile, we notice that when the LED is turned on, run durations are longer in every direction. This is consistent with our previous results obtained in isotropic conditions, where we have shown that additional intensity tends to increase the run durations. As this plot is isotropic, run durations cannot explain the bias in the motility.

To push further the investigation in the origin of the observed bias in motility, we plot in FIGURE 8(b) the run frequencies (i.e the percentages of run events in the considered angular segment) as a function of the angular direction. It defines an angular histogram of the runs' occurrences. When the LED is turned off, we obtain again circles, showing no preferential direction for the runs. However, when it is turned on, the proportion of runs is distorted, with higher probabilities for $\theta$ close to 0. This infers that most of the run periods result in displacements toward the light source. The bias of the motility is then due to the higher number of runs in the direction of the light rather than longer runs.

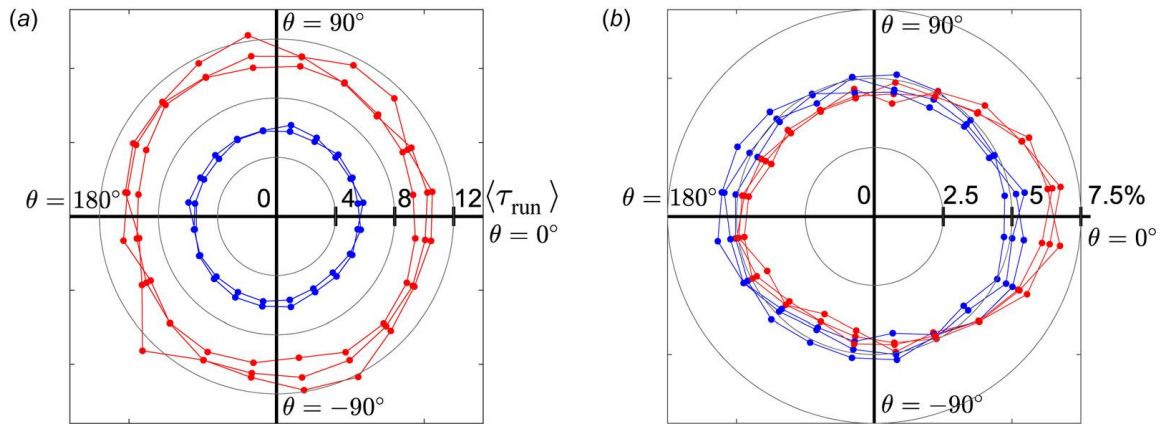

**FIGURE 8:** Average run (a) durations and (b) frequencies as a function of their angular directions, $\theta = 0°$ being the direction towards the light source. Red: LED is turned on, blue: LED is turned off.

**Discussion**

The way microorganisms face changes in and adapt to their environment is still not well understood. Our experimental setup enables us to get access to the dynamic response of *Synechocystis* undergoing abrupt light changes. Two hypothesis describing the ability to react to an external perturbation exist: (i) bacteria integrate on their displacement the external conditions, and eventually choose the one which seems to be the most favorable for their development, or (ii) bacteria "feel" immediately the direction of the perturbation and respond to it [7].

The first hypothesis is verified for instance with chemotaxis of *E.Coli* bacteria, where run periods are longer in the direction of the nutrients [1]. We show that the response of *Synechocystis* to light intensity steps display similarities with that of *E.Coli* submitted to chemoattractant concentration steps. The common shape of their run durations facing two different kinds of perturbations raises the question of the ubiquity of the bacterial response to different kinds of stresses.

However, for our study on the phototactic response, the fact that the bias emerges as soon as the LED is turned on suggests that the bacteria detect the light direction as soon as it appears. This feature, also observed in other studies [4,21] is consistent with the description of Schuergers et al. [22], where cell membrane acts as a micro-lens and focuses the light flux to trigger the formation or dissolution of the pili responsible for the bacterial displacement.

We also show that, for *Synechocystis* cells, run durations toward the light source are not longer, which is also in agreement with the second hypothesis of direct recognition of light direction mentioned above. Indeed, this observation is different from the mechanism of chemotaxis of *E.Coli*, where runs are longer when they point to the nutrient source. Here, we observe an increase of the run durations in every direction with additional light intensity; consistent with the study we have carried out in isotropic conditions; while runs occur more frequently toward the light source. Hence, run durations seem to respond to light intensity and run directions respond to light direction.

This can suggest that the mechanisms that trigger run durations and the ones that govern run directions do not follow the same pathway in the *Synechocystis'* metabolism. Run durations seem to be more susceptible to light intensity, whereas their directions are governed by the orientation of the incident light flux.

**Conclusions**

We investigate in this paper the response of *Synechocystis* cells to light perturbations aimed at active control of the diffusion coefficient of this microorganism in biofilm. The photosynthetic micro-organisms are sensitive to light intensity and can adapt their diffusion by triggering the characteristic times of their intermittent motility. With higher illumination, bacteria have a greater propensity to be in a run mode during which they perform longer displacements and therefore increase their diffusion coefficient.

We model analytically the response of *Synechocystis* cells by using the linear response theory. A response function that has been proposed to describe *E.Coli* chemotaxis was found suitable to fit our experimental data, with an adaptation of the numerical values of the response function parameters.

Under directional light flux, *Synehcocystis* cells perform a phototactic motility and head toward the light source. This biased motility stems from the averaged displacements during run periods, which is no longer random. We show that the bias is the result of the number of runs, which is greater toward the light source, and not of longer runs in this direction. Brought together, these results suggest distinct pathways for the recognition of light intensity and of its direction in this prokaryote micro-organism, that can be used in the active control of bacterial flows.

**ACKNOWLEDGEMENTS**

We would like to thank Annick Méjean, who provided us the bacterial strain used in this study, and advised us for the culture process. We also would like to thank Jean Pierre Thermeau and Mojtaba Jarrahi for the fruitful discussions we had during this study.

**REFERENCES**

[1]  Berg, H. C., and Brown, D. a, 1972, "Chemotaxis in Escherichia Coli Analysed by Three-Dimensional Tracking.," Nature, **239**(5374), pp. 500–504.

[2]  Segall, J. E., Block, S. M., and Berg, H. C., 1986, "Temporal Comparisons in Bacterial Chemotaxis," PNAS, **83**(December), pp. 8987–8991.

[3]  Yoshihara, S., and Ikeuchi, M., 2004, "Phototactic Motility in the Unicellular Cyanobacterium Synechocystis Sp. PCC 6803," Photochem. Photobiol. Sci., **3**(6), p. 512.

[4]  Ursell, T., Chau, R. M. W., Wisen, S., Bhaya, D., and Huang, K. C., 2013, "Motility Enhancement through Surface Modification Is Sufficient for Cyanobacterial Community Organization during Phototaxis," PLoS Comput. Biol., **9**(9).

[5]  Wilde, A., and Mullineaux, C. W., 2017, "Light-Controlled Motility in Prokaryotes and the Problem of Directional Light Perception," FEMS Microbiol. Rev., **41**(6), pp. 900–922.

[6]  Schuergers, N., Mullineaux, C. W., and Wilde, A., 2017, "Cyanobacteria in Motion," Curr. Opin. Plant Biol., **37**, pp. 109–115.

[7]  Chau, R. M. W., Bhaya, D., and Huanga, K. C., 2017, "Emergent Phototactic Responses of Cyanobacteria under Complex Light Regimes," MBio, **8**(2), pp. 1–15.

[8]  Kaneko, T., Sato, S., Kotani, H., Tanaka, A., Asamizu, E., Nakamura, Y., Miyajima, N., Hirosawa, M., Sugiura, M., Sasamoto, S., Kimura, T., Hosouchi, T., Matsuno, A., Muraki, A., Nakazaki, N., Naruo, K., Okumura, S., Shimpo, S., Takeuchi, C., Wada,

T., Watanabe, A., Yamada, M., Yasuda, M., and Tabata, S., 1996, "Sequence Analysis of the Genome of the Unicellular Cyanobacterium Synechocystis Sp. Strain PCC6803. II. Sequence Determination of the Entire Genome and Assignment of Potential Protein-Coding Regions," DNA Res., **3**(3), pp. 109–136.

[9] Ikeuchi, M., and Tabata, S., 2001, "Synechocystis Sp. PCC 6803 – a Useful Tool in the Study of the Genetics of Cyanobacteria," Photosynth. Res., **70**, pp. 73–83.

[10] Vourc'h, T., Léopoldès, J., Méjean, A., and Peerhossaini, H., 2016, "Motion of Active Fluids: Diffusion Dynamics of Cyanobacteria," *Fluids Engineering Division Summer Meeting*, ASME.

[11] Vourc'h, T., Peerhossaini, H., Léopoldès, J., Méjean, A., Cassier-Chauvat, C., and Chauvat, F., 2018, "Slowdown of the Surface Diffusion during Early Stages of Bacterial Colonization," Phys. Rev. E, **97**(3), pp. 17–20.

[12] Fadlallah, H., Jarrahi, M., Herbert, É., Ferrari, R., Méjean, A., and Peerhossaini, H., 2020, "Active Fluids : Effects of Hydrodynamic Stress on Growth of Self-Propelled Fluid Particles," J. Appl. Fluid Mech., **13**(2).

[13] De Gennes, P.-G., 2004, "Chemotaxis : The Role of Internal Delays," Eur. Biophys. J., **33**(8), pp. 691–693.

[14] Clark, D. A., and Grant, L. C., 2005, "The Bacterial Chemotactic Response Reflects a Compromise between Transient and Steady-State Behavior.," Proc. Natl. Acad. Sci. U. S. A., **102**(26), pp. 9150–5.

[15] Celani, A., and Vergassola, M., 2010, "Bacterial Strategies for Chemotaxis Response," Proc. Natl. Acad. Sci., **107**(4), pp. 1391–1396.

[16] Berg, H., Darnton, N., and Jaffe, N., "Object Tracking Software."

[17] Zaburdaev, V., Biais, N., Schmiedeberg, M., Eriksson, J., Jonsson, A. B., Sheetz, M. P., and Weitz, D. A., 2014, "Uncovering the Mechanism of Trapping and Cell Orientation during Neisseria Gonorrhoeae Twitching Motility," Biophys. J., **107**(7), pp. 1523–1531.

[18] Segall, J. E., Block, S. M., and Berg, H. C., 1986, "Temporal Comparisons in Bacterial Chemotaxis," PNAS, **83**, pp. 8987–8991.

[19] Clark, D. A., and Grant, L. C., 2005, "The Bacterial Chemotactic Response Reflects a Compromise between Transient and Steady-State Behavior.," Proc. Natl. Acad. Sci. U. S. A., **102**(26), pp. 9150–9155.

[20] Ng, W., Grossman, A. R., and Bhaya, D., 2003, "Multiple Light Inputs Control Phototaxis in Multiple Light Inputs Control Phototaxis in Synechocystis Sp. Strain PCC6803," **185**(5), pp. 1599–1607.

[21] Nakane, D., and Nishizaka, T., 2017, "Asymmetric Distribution of Type IV Pili Triggered by Directional Light in Unicellular Cyanobacteria," Proc. Natl. Acad. Sci., **114**(25), pp. 6593–6598.

[22] Schuergers, N., Lenn, T., Kampmann, R., Meissner, M. V., Esteves, T., Temerinac-Ott, M., Korvink, J. G., Lowe, A. R., Mullineaux, C. W., and Wilde, A., 2016, "Cyanobacteria Use Micro-Optics to Sense Light Direction," Elife, **5**, pp. 1–16.

**Figure Captions**

| | |
|---|---|
| Fig. 1 | Schematic view of the experimental setups used in this study. (a) Response to an isotropic perturbation, (b) response to an anisotropic (directional) perturbation. |
| Fig. 2 | Temporal profile of the light intensity used in the isotropic experiment. During Phase I, intensity is set to $I_0$ for two hours. It is then raised to $I_0+\Delta I$ for one hour (Phase II), and set to $I_0$ again for the subsequent hour (Phase III). The latter cycle is repeated one (Phases IV-V). |
| Fig. 3 | Temporal evolution of the diffusion coefficient normalized by their final value. Plain line: experimental results, and dotted line: analytical formula (EQ. 13). Colored areas indicate +/- standard deviation. |
| Fig. 4 | Trajectory of a bacterial cell (668 s). Runs correspond to red lines and tumble to blue dots. For the phototaxis assay, we define the angle $\theta$ between the direction of the run and the direction towards the light source, which is indicated by the arrow. |
| Fig. 5 | Temporal evolution of the inverse (a) run and (b) tumble times, rescaled by their final value. Black Plain line: model described by EQ. (7-8). Colored areas indicate +/- standard deviation. |
| Fig. 6 | Average velocity along the longitudinal and transverse directions as a function of the time elapsed since the conditions have changed. Red: After |

the LED has been turned on (anisotropic illumination) and blue: after the LED has been turned off (isotropic illumination). Black dashed line indicates a zero velocity corresponding to an absence of biased movement. Colored areas indicate +/- standard deviation.

Fig. 7 Temporal evolution of the averaged projection during runs periods along longitudinal $x$ and transverse $y$ directions. Plain black line shows the state of the LED (turned off when it is 0, turned on elsewhere). Colored areas indicate +/- standard error.

Fig.8 Average run (a) durations and (b) frequencies as a function of their angular directions, $\theta = 0°$ being the direction towards the light source. Red: LED is turned on, blue: LED is turned off.

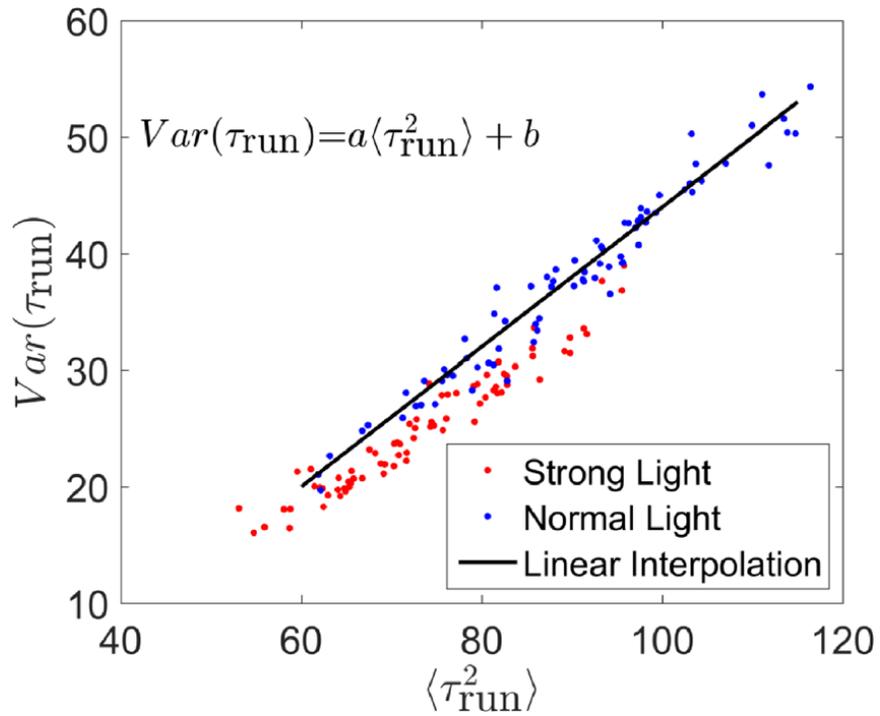

**APPENDIX A:** Variance of the run time distribution as a function of the mean squared run times, for the two different light intensities used in the isotropic study. Black line: linear empirical fit used for the derivation of EQ. (13).